\documentclass[aip,jcp,reprint,amsmath,amssymb,floatfix,citeautoscript]{revtex4-1}
\usepackage{amsmath}
\usepackage[usenames,dvipsnames]{color}
\usepackage{amssymb}
\usepackage{braket}
\usepackage{graphicx}
\usepackage{ragged2e}
\usepackage[labelsep=period,format=plain,justification=justified,font=footnotesize]{caption}
\DeclareCaptionJustification{myjust}{\justifying}
\usepackage{times}
\usepackage{txfonts}

\begin{document}

\title{Approximate but accurate quantum dynamics from the Mori formalism: II. Equilibrium correlation functions}
\author{Andr\'{e}s Montoya-Castillo\footnote{Corresponding Author}}
\email{am3720@columbia.edu}
\affiliation{Department of Chemistry, Columbia University, New York, New York, 10027, USA}
\author{David R. Reichman}
\email{drr2103@columbia.edu}
\affiliation{Department of Chemistry, Columbia University, New York, New York, 10027, USA}

\date{\today}

\begin{abstract}
 The ability to efficiently and accurately calculate equilibrium time correlation functions of many-body condensed phase quantum systems is one of the outstanding problems in theoretical chemistry.  The Nakajima-Zwanzig-Mori formalism coupled to the self-consistent solution of the memory kernel has recently proven to be highly successful for the computation of nonequilibrium dynamical averages. Here, we extend this formalism to treat symmetrized equilibrium time correlation functions for the spin-boson model. Following the first paper in this series [A. Montoya-Castillo and D. R. Reichman, J. Chem. Phys. \textbf{144}, 184104 (2016)], we use a Dyson-type expansion of the projected propagator to obtain a self-consistent solution for the memory kernel that requires only the calculation of normally evolved auxiliary kernels. We employ the approximate mean-field Ehrenfest method to demonstrate the feasibility of this approach.  Via comparison with numerically exact results for the correlation function $\mathcal{C}_{zz}(t) = \mathrm{Re}\langle \sigma_z(0)\sigma_z(t)\rangle$, we show that the current scheme affords remarkable boosts accuracy and efficiency over bare Ehrenfest dynamics. We further explore the sensitivity of the resulting dynamics to the type of kernel closures and the accuracy of the initial canonical density operator.  
\end{abstract}

\maketitle


\normalsize

\section{Introduction}

By encoding a system's response to a weak perturbation \cite{KuboStatMechII}, equilibrium time correlation functions (ECF) provide direct access to important dynamical quantities such as transport coefficients,\cite{Kubo1957a, Kubo1957b} absorption spectra \cite{PrinciplesNonlinearSpec}, and chemical rate constants \cite{Kubo1957b, Yamamoto1960, Voth1989a}. It is therefore not surprising that the development of accurate and efficient approaches to the calculation of ECFs has been a focus of intense theoretical investigation, spawning a rich array of analytical and computational methods.

Attempts at calculating ECFs face the challenge of striking a balance between computational tractability and accuracy.  On the more computationally expensive side lie numerically exact schemes, which are generally restricted to small, idealized models and tend to scale unfavorably, and at worst exponentially, with simulation time \cite{Mak1990, Mak1991, Egger1992, Egger1994a, Shao2001, Shao2002, Tanimura2014, Tanimura2015, Song2015}.  Approximate methods, on the other hand, are more scalable to realistic, multidimensional systems, but suffer from limited accuracy.  These include perturbative approaches \cite{Redfield1965, Bloch1957, Leggett1987, Aslangul1986a}, analytic continuation \cite{Jarrell1996,Gallicchio1996, Boninsegni1996, Sim2001, Krilov2001, Golosov2003}, quantum mode coupling theory \cite{Rabani2002, Reichman2002, Rabani2004,Rabani2005,Markland2011}, and quasi- \cite{Gerber1982, Stock1995, Tully1971,Tully1990,Tully1998a} and semi-classical \cite{Meyer1979, Stock1997, Wang1998b, Sun1998, Shi2003d, Miller1998, Miller2001a,Miller2009, Liu2007, Cao1994b, Jang1999a, Jang1999, Craig2004, Craig2005, Craig2005b, Habershon2013, Egorov1998, Egorov1999,Shi2003c} schemes.  While each method has its virtues, all suffer from limited applicability, either due the violation of parameter regime restrictions, the breakdown of uncontrolled approximations, or convergence problems.  As a result, despite the availability of many methods for the calculation of ECFs, there remains a clear need for the development of widely applicable, accurate, and computationally efficient approaches.   

A successful marriage of computational efficiency and accuracy may be enabled through a judicious use of the Mori formalism. Based on the projection operator technique \cite{Grabert1982, FickSauermann}, the Mori approach provides a simple, low-dimensional equation of motion for the ECF, called a generalized quantum master equation (GQME).  The reduced dimensionality of the GQME corresponds to that of the space spanned by the observables probed in the ECF, while the influence of the excluded degrees of freedom is encoded in the memory kernel. Calculation of the memory kernel, in turn, is beset by two difficulties: application of the projected propagator, $e^{i\mathcal{Q}\mathcal{L}t}$, and the full dimensionality of the original problem. The former can be sidestepped using a Dyson-type expansion, which leads to a self-consistent equation for the memory kernel that requires only the calculation of projection-free auxiliary kernels \cite{Shi2003}.  The latter, however, continues to plague the calculation of the auxiliary kernels.  Nevertheless, this approach has been vigorously pursued to obtain nonequilibrum averages in the context of impurity-type models where the auxiliary kernels have been obtained via numerically exact \cite{Shi2003, Zhang2006a, Cohen2011, Cohen2013, Cohen2013a, Wilner2013, Wilner2014} or approximate methods \cite{Shi2004a, Kelly2013, Kelly2015, Pfalzgraff2015, Montoya2016a, KellyMontoya2016}.  These studies have shown that the short lifetime of the memory kernels can lead to dramatic increases in computational efficiency, regardless of the method used to compute the memory kernels \cite{Shi2003, Zhang2006a, Cohen2011, Cohen2013, Cohen2013a, Wilner2013, Wilner2014, Shi2004a, Kelly2013, Kelly2015, Pfalzgraff2015, Montoya2016a, KellyMontoya2016}.  Just as importantly, when approximate methods are used, the Mori approach has also provided impressive boosts in accuracy over the bare approximate dynamics \cite{Shi2004a, Kelly2013, Kelly2015, Pfalzgraff2015, Montoya2016a, KellyMontoya2016}.  Hence, the approach based on the GQME coupled to the self-consistent solution of the memory kernel shows great promise as a means of increasing the efficiency and, when appropriate, accuracy of dynamical methods.

Here we argue that the remarkable boosts in efficiency and accuracy afforded by the GQME approach can be extended to arbitrary systems and dynamical quantities beyond simple nonequilibrium averages.  To show the viability of the approach, we specialize the Mori treatment to the symmetrized ECFs for the spin variables of the spin-boson (SB) model \cite{Leggett1987, Weiss}. In the same spirit as the first paper of this series \cite{Montoya2016a}, we calculate the auxiliary kernels necessary for the self-consistent solution of the memory kernel via the mean-field Ehrenfest method and assess the potential benefits of this approach in terms of increases in efficiency and accuracy (we henceforth refer to the this framework as the GQME+MFT approach). In this work, we also endeavor to elucidate the dependence of the GQME dynamics on the choice of closure, and try to provide further evidence for the claim that the source of the improvement over bare semiclassical dynamics afforded by the GQME framework depends, at least partially, on the \textit{exact} sampling of \textit{distinct} initial conditions necessary in the calculation of the auxiliary kernels. It also bears remarking that the Mori approach can be easily generalized to multi-time correlation functions and is applicable to a wide variety of systems.  In fact, a major advantage of the Mori formulation is that it can naturally address problems where the system-bath dichotomy is absent, such as spin and fermion lattice models \cite{Rasetti1991, Sachdev2011, Giamarchi2004}, and quantum fluids \cite{Feenberg1969, PinesNozieres1999, Poulsen2005, Markland2011}.

The paper is organized as follows.  In Sec.~\ref{Ch4Sec:MoriApproach}, we briefly introduce the SB model and the projection operator that allows for the investigation of symmetrized ECFs of the Pauli matrices.  Sec.~\ref{Ch4Subsec:EffieciencyAccuracy}, compares the dynamics obtained via the GQME+MFT approach to numerically exact results for the symmetrized spin-spin correlation function $\mathcal{C}_{zz}(t) = \mathrm{Re} \langle \sigma_z(0) \sigma_z(t)\rangle$.  Sec.~\ref{Ch4Subsec:Closures} explores the dependence of GQME results on the choice of closure and the accuracy of the initial conditions. Sec.~\ref{Ch4Sec:Conclusions} is devoted to our concluding remarks.

\section{Mori Approach}
\label{Ch4Sec:MoriApproach}

As in paper I, we apply the Mori formalism for ECFs to the SB model, which is representative of typical condensed phase systems that exhibit nontrivial decoherence and dissipation patterns \cite{Leggett1987, Weiss}. We note, however, that the Mori approach is general and may be applied to any Hamiltonian system, including those where the system-bath distinction is absent.  This point is of crucial importance for the modeling of quantum ECFs such as those associated with absorption spectra in liquids \cite{PinesNozieres1999}.

The SB Hamiltonian takes the form $H = H_S + H_B + H_{SB}$, where 
	\begin{equation}
	H_S = \varepsilon\sigma_z + \Delta \sigma_x,
	\end{equation}	 
	corresponds to the system part of the Hamiltonian.  Here, $2 \varepsilon$ corresponds the energy difference between the two sites, $\Delta$, which is assumed to be static, represents the tunneling matrix element, and $\sigma_i$ corresponds to the $i^{th}$ Pauli matrix. 
	
	The bath Hamiltonian consists of independent harmonic oscillators, 	\begin{equation}
	H_B = \frac{1}{2} \sum_k \Big[\hat{P}_k^2 + \omega_k^2 \hat{Q}_k^2\Big],
	\end{equation}
	where $P_k$, $Q_k$ and $\omega_k$ are the mass-weighted momenta, coordinates, and the frequency for the $k^{th}$ harmonic oscillator, respectively. 
	The system-bath coupling is assumed to be of the form, 
	\begin{equation}\label{Ch4Eq:SBInteraction}
	V = \alpha \sigma_z \sum_k c_k \hat{Q}_k,
	\end{equation}
	where $c_k$ is the coupling constant describing the strength of the interaction between the system and the $k^{th}$ oscillator and $\alpha = \pm 1$.  The system-bath interaction is fully characterized by the spectral density, 
	\begin{equation}
	\begin{split}
	J(\omega) &= \frac{\pi}{2}\sum_k \frac{c_k^2}{\omega_k}\delta(\omega - \omega_k)\\
	&= \xi \omega e^{-\omega / \omega_c} \label{Ch4Eq:OhmicSD},
	\end{split}
	\end{equation}
which encodes the frequency-resolved coupling between the system and the oscillators that compose the bath.  The second line in Eq.~(\ref{Ch4Eq:OhmicSD}) corresponds to the often used Ohmic form for the spectral density \cite{Leggett1987} with an exponential cutoff.  Here, $\omega_c$ is the cutoff frequency, which determines the correlation time for the bath at finite temperature.  The Kondo parameter, $\xi$, describes the strength of the system-bath coupling and is proportional to the reorganization energy, $\lambda  = \xi \omega_c / \pi = \pi^{-1} \int_0^{\infty} d\omega\ J(\omega)/\omega$, which represents energy dissipated after the system makes a Frank-Condon transition. We also note that the present approach is neither limited to the SB model nor to the Ohmic form of the spectral density.

\subsection{Mori-type GQME}
\label{Ch4Subsec:GeneralizedNZMEq}
As established in Paper I, the Mori (as well as the Nakajima-Zwanzig) approach  stems from the generalized Mori-Zwanzig-Nakajima equation of motion for the propagator, 
	\begin{equation} \label{Ch4Eq:GeneralNZMEq}
	\begin{split}
	\frac{d}{dt}e^{i\mathcal{L}t} &= ie^{i\mathcal{L}t}\mathcal{P}\mathcal{L} + i \mathcal{Q}e^{i\mathcal{L}\mathcal{Q}t}\mathcal{L} \\
	&\qquad -  \int_0^{t}d\tau\ e^{i\mathcal{L}(t-\tau)}\mathcal{P}\mathcal{L}\mathcal{Q}e^{i\mathcal{L}\mathcal{Q}\tau}\mathcal{L},
	\end{split}
	\end{equation}
where $\mathcal{P}$ is the projection operator, which consists of the dynamical operators whose correlations we seek, and $\mathcal{Q} = 1 - \mathcal{P}$ is the complementary projection operator. 

For ECFs, the Mori-type projector commonly takes the form, 
	\begin{equation}\label{Ch4Eq:ProjectionOperator}
	\mathcal{P} = \frac{|\mathbf{A})(\mathbf{A}|}{(\mathbf{A}|\mathbf{A})},
	\end{equation}
where the vector $\mathbf{A}$ consists of an operator $A$ and its time derivative $\dot{A} = i\mathcal{L}A$, which ensures the idempotency of the projection operator \cite{Mori1965, Reichman2005, Rabani2005, KellyMontoya2016}. Instead of taking this approach, we employ a projector that recovers the symmetrized spin ECFs for the SB model. Hence, we take $\mathbf{A}$ to consist of the Pauli matrices, $\sigma_i$, where $i \in \{ x,y,z \}$, and define the inner product as, 
	\begin{equation}\label{Ch4Eq:InnerProduct}
	(\boldsymbol{\sigma}|\mathcal{O}|\boldsymbol{\sigma})_{nm} \equiv \frac{1}{2} \mathrm{Tr}\Big[ \rho \{ \sigma_n, (\mathcal{O}\sigma_m) \} \Big],
	\end{equation}
	where $\{A, B\} = AB + BA$ is the anticommutator, $\rho = e^{-\beta H}/\mathrm{Tr}[e^{-\beta H}]$ is the canonical density operator, and $\beta = [k_BT]^{-1}$ is the inverse thermal energy.  Using the components of the projection operator to close Eq.~(\ref{Ch4Eq:GeneralNZMEq}) from both sides, we obtain the Mori-type GQME, 
	\begin{equation}\label{Ch4Eq:GQME}
	\frac{d}{dt}\mathcal{C}(t) = \mathcal{C}(t)\dot{\mathcal{C}}(0) -  \int_0^{t}d\tau\ \mathcal{C}(t-\tau)\mathcal{K}(\tau), 
	\end{equation}
	where $\mathcal{C}_{nm}(t) = \mathrm{Tr}[ \rho \{ \sigma_n, \sigma_m(t) \} ]/2 \equiv \mathrm{Re} \langle \sigma_n(0)\sigma_m(t)\rangle$ is the symmetrized or real part of the correlation function for the Pauli spin operators.  The memory kernel takes the following form
	\begin{align}
	\mathcal{K}(t) &= (\boldsymbol{\sigma}| \mathcal{L}\mathcal{Q}e^{i\mathcal{Q}\mathcal{L}t}\mathcal{Q}\mathcal{L} |\boldsymbol{\sigma}). \label{Ch4Eq:MemoryKernel}
	\end{align}
Given the difficulties associated with treating the dynamics required by the projected propagator in Eq.~(\ref{Ch4Eq:MemoryKernel}), we follow Shi and Geva \cite{Shi2003} and employ the Dyson decomposition, 
	\begin{align}
	e^{(A+B)t} &= e^{At} + \int_0^t ds\  e^{As}Be^{(A+B)(t - s)} \label{Ch4Eq:DysonEq1}\\
	&= e^{At} + \int_0^t ds\  e^{(A+B)s}Be^{A(t - s)} \label{Ch4Eq:DysonEq2}
	\end{align}
to obtain the $\mathcal{Q}$-forward ($f$) and $\mathcal{Q}$-backward ($b$) self-consistent expansions of the memory kernel,
	\begin{align}
	\mathcal{K}(t) &= \mathcal{K}^{(1)}(t) + \int_0^t d\tau\ \mathcal{K}(t - \tau)\mathcal{K}^{(3f)}(\tau), \label{Ch4Eq:SelfConsistentK_forward}\\
	&= \mathcal{K}^{(1)}(t) + \int_0^t d\tau\ \mathcal{K}^{(3b)}(t-\tau)\mathcal{K}(\tau), \label{Ch4Eq:SelfConsistentK_backward} 
	\end{align} 
where the normally evolved auxiliary kernels 
	\begin{align}
	\mathcal{K}^{(1)}(t) &= (\boldsymbol{\sigma}| \mathcal{L}\mathcal{Q}e^{i\mathcal{L}t}\mathcal{Q}\mathcal{L} |\boldsymbol{\sigma}) , \label{Ch4Eq:K1}\\
	\mathcal{K}^{(3f)}(t) &= -i\ (\boldsymbol{\sigma}| e^{i\mathcal{L}t} \mathcal{Q} \mathcal{L} |\boldsymbol{\sigma}) \label{Ch4Eq:K3f}, \\
	\mathcal{K}^{(3b)}(t) &= -i\ (\boldsymbol{\sigma}| \mathcal{L}\mathcal{Q}e^{i\mathcal{L}t}  |\boldsymbol{\sigma}) \label{Ch4Eq:K3b}, 
	\end{align}
	can be obtained via direct simulation.  For more details about the derivation, we refer the reader to Ref. \onlinecite{Montoya2016a}.
	
As discussed in the first paper of this series \cite{Montoya2016a}, different closures may lead to different results when using approximate dynamics to calculate the auxiliary kernels.  For that reason we also explore the effect of the three additional closures that selectively replace the action of the Liouvillian with time-derivatives.  For completeness, these alternative closures are reproduced below. In the first set, we replace the action of the Liouvillian acting on operators that require dynamic sampling with the time-derivative, such that 
	\begin{align}
	\mathcal{K}^{(1)}_{1}(t) &= \dot{\mathcal{K}}^{(3b)}(t) -  \mathcal{K}^{(3b)}(t) \dot{\mathcal{C}}(0), \label{Ch4Eq:K1b1ActDeriv}\\
	\mathcal{K}^{(3b)}_{1}(t) &=  \mathcal{K}^{(3b)}(t)\label{Ch4Eq:K3b1ActDeriv},\\
	\mathcal{K}^{(3f)}_{1}(t) &= -\dot{\mathcal{C}}(t) +  \mathcal{C}(t)\dot{\mathcal{C}}(0). \label{Ch4Eq:K3f1ActDeriv}
	\end{align}

The second type replaces the action of the Liouvillian on static operators with the time derivative, 
	\begin{align}
	\mathcal{K}^{(1)}_{2}(t) &= \dot{\mathcal{K}}^{(3f)}(t) - \dot{\mathcal{C}}(0)\mathcal{K}^{(3f)}(t) , \label{Ch4Eq:K1b1PasDeriv}\\
	\mathcal{K}^{(3b)}_{2}(t) &= -\dot{\mathcal{C}}(t)+ \dot{\mathcal{C}}(0)\mathcal{C}(t)\label{Ch4Eq:K3b1PasDeriv},\\
	\mathcal{K}^{(3f)}_{2}(t) &=  \mathcal{K}^{(3f)}(t). \label{Ch4Eq:K3f1PasDeriv}
	\end{align}
	
The final type replaces all Liouvillian operators with time derivatives, 
	\begin{align}
	\mathcal{K}^{(1)}_{3}(t) &= -\ddot{\mathcal{C}}(t) + \{ \dot{\mathcal{C}}(t), \dot{\mathcal{C}}(0) \} -  \dot{\mathcal{C}}(0)\mathcal{C}(t)\dot{\mathcal{C}}(0), \label{Ch4Eq:K1b1DDeriv}\\
	\mathcal{K}^{(3b)}_{3}(t) &=  \mathcal{K}^{(3b)}_2(t)\label{Ch4Eq:K3b1DDeriv},\\
	\mathcal{K}^{(3f)}_{3}(t) &= \mathcal{K}^{(3f)}_1(t). \label{Ch4Eq:K3f1DDeriv}
	\end{align}
	
It bears repeating that \textit{all} closures presented above are permitted by quantum mechanics and, when exact methods are used to calculate the auxiliary kernels, the memory kernels and GQME dynamics produced will be equivalent, regardless of the closure.  However, as was shown in the nonequilibrium case \cite{Montoya2016a}, the memory kernels and GQME dynamics obtained from auxiliary kernels calculated via approximate schemes can differ from each other, depending on the closure. Explicit expressions for the various closures in the context of the SB model can be found in Appendix \ref{Ch4App:AuxKernels}.

\section{Results}
\label{Ch4Sec:Results}
		
As in the first paper of this series \cite{Montoya2016a}, we employ the Ehrenfest method to calculate the auxiliary kernels necessary for the extraction of the memory kernels. The Ehrenfest approach is a simple quasi-classical scheme where the system (bath) evolves in the mean field of the bath (system).  While most implementations of the Ehrenfest method (and other low-level quasi- and semi-classical theories such as the surface hopping \cite{Tully1971, Tully1990} and linearized semiclassical initial value representation schemes \cite{Wang1998b, Sun1998, Shi2003d} (LSC-IVR) use approximate forms for the fully correlated Boltzmann factor \cite{Wang1998b, Poulsen2003, Liu2008a, Liu2009}, we employ a numerically exact representation which relies on the path integral approach developed in Ref.~\onlinecite{Montoya2016c}. The importance of the exact rendering of the canonical density operator will become evident in Sec. \ref{Ch4Subsec:EffieciencyAccuracy}. Within this path integral framework, it is possible to produce expressions for the canonical distribution that increase in accuracy with the number of path integral slices, $N$, used. The number of path integral steps taken in the rendering of a given realization of $\rho$ is determined by the number of slices necessary for the convergence of static and dynamic data as a function of $N$. Details regarding the path integral approach to the canonical density are included in Appendix \ref{Ch4App:PIrho}, and those regarding the implementation of the Ehrenfest method in Appendix \ref{Ch4App:AuxKernels}.

The protocol for the extraction of $\mathcal{K}(t)$ can be summarized as follows.  Using the approximate auxiliary kernels obtained via the Ehrenfest scheme, we solve Eqs.~(\ref{Ch4Eq:SelfConsistentK_forward}) and (\ref{Ch4Eq:SelfConsistentK_backward}) iteratively until the relative error $R.E.$ is negligible, i.e., $R.E. < 10^{-10}$, where $R.E. = \mathrm{max}[\mathrm{abs}[\mathcal{K}_{n+1}(t) - \mathcal{K}_n(t)]]$ is the maximum absolute difference between two subsequent iterations of the memory kernel.  With a converged solution for the memory kernel, we numerically integrate  Eq.~(\ref{Ch4Eq:GQME}) using a second-order Runge-Kutta procedure, subject to the appropriate initial conditions, namely $\mathcal{C}_{ii}(t =0 ) = 1$.  To assess the viability of the GQME+MFT procedure in the equilibrium context we compare the solution of Eq.~(\ref{Ch4Eq:GQME}) using the self-consistently extracted $\mathcal{K}(t)$ to numerically exact results for $\mathcal{C}_{zz}(t)$ for the SB model with an Ohmic spectral density. We further specify that, in the same vein as previous implementations of the GQME coupled to the self-consistent solution of the memory kernel \cite{Shi2003, Shi2004a, Zhang2006a, Kelly2013, Kelly2015, Pfalzgraff2015, Cohen2011, Cohen2013, Cohen2013a, Wilner2013, Wilner2014, Montoya2016a, KellyMontoya2016}, the finite lifetime of the memory kernel is explicitly included in the GQME evolution algorithm as a cutoff time, $\tau_c$, beyond which the memory kernel is set to zero.  In practice, $\tau_c$ is identified as any time point in the stability plateau corresponding to the range of time during which the memory kernel ceases to influence the GQME dynamics. When numerically exact methods are used to obtain the auxiliary kernels, this stability plateau is infinitely long.\footnote{It should be noted that there are problems where the memory kernels never decays to zero. A clear example of this is the case of a SB model where the bath consists of \textit{one} oscillator, where the persistent quantum coherence results in a memory kernel that never decays. For problems where decoherence and dissipation are expected to be significant, a non-decaying memory kernel is a sign that the projection operator has to be chosen differently. Clearly, the GQME formalism cannot provide any advantages with respect to computational efficiency for cases where the memory kernel never decays. }  Conversely, when approximate methods are used instead, the limited accuracy of the underlying method may cause the stability plateau to be finite or, in extreme cases, nonexistent.  In all cases studied in this work, a well-defined stability plateau exists and all cutoff times for the memory kernels are specified (See Sec.~\ref{Ch4Subsec:LongTimeDynamics} for a more detailed exploration of how to determine the plateau of stability for difficult cases). 

\subsection{Improvements in Efficiency and Accuracy}
\label{Ch4Subsec:EffieciencyAccuracy}

For nonequilibrium averages, the GQME+semiclassics approach has been shown to be uniformly beneficial in reducing the cost of the dynamics via short-lived memory kernels and generally advatageous in correcting approximate dynamics, especially for biased systems coupled weakly to the bath \cite{Shi2003,Zhang2006a,Cohen2011,Cohen2013,Cohen2013a,Wilner2013, Wilner2014,Shi2004a,Kelly2013,Kelly2015,Pfalzgraff2015,Montoya2016a,KellyMontoya2016}.   In the following, we show that similar benefits can be reaped in the equilibrium case.  

\begin{figure}
\centering
\includegraphics[width=8.5cm]{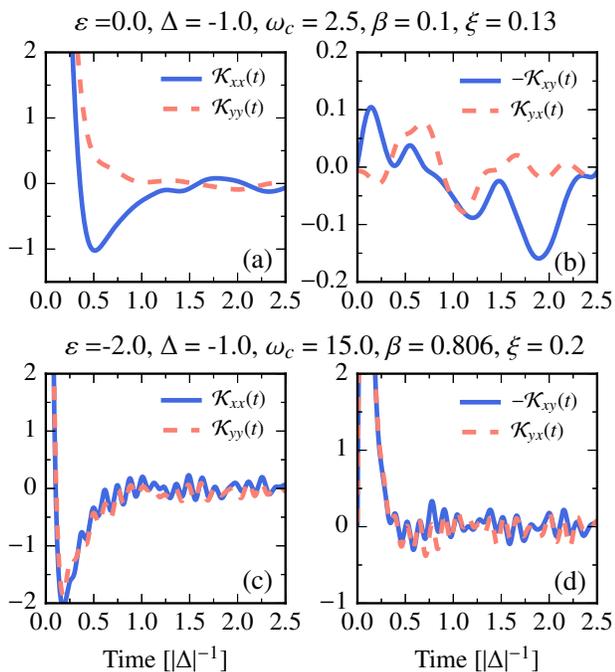} 
\caption{Memory kernels obtained using the $cf0$ closure for two realizations of the SB model.  For panels (a)--(b), $\alpha = 1$ and for (c)--(d), $\alpha = -1$. By symmetry requirements, all other components of the memory kernel are identically zero.}\label{Ch4Fig:K_cf0}
\end{figure}

\begin{figure}[t]
\centering
\includegraphics[width=8.5cm]{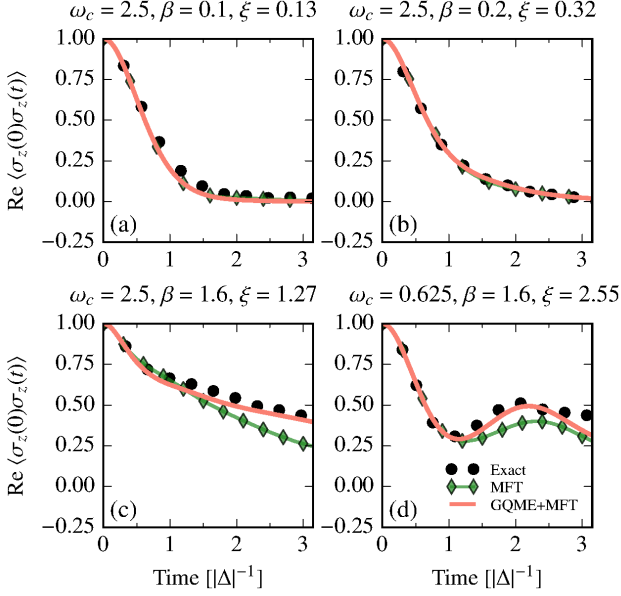} 
\caption{Comparison of $\mathcal{C}_{zz}(t) = \mathrm{Re}\ \langle \sigma_z(0)\sigma_z(t)\rangle$ obtained from the $cf0$ memory kernels for the unbiased SB model ($\varepsilon = 0$) with $\Delta = -1$ and $\alpha = 1$.  Converged representations of the canonical density required $N = 0,1,5,6$ path integral slices for panels (a)--(d), respectively. Exact results are obtained from Ref.~\onlinecite{Mak1991}.}\label{Ch4Fig:Czz_MC}
\end{figure}
\begin{figure}[t]
\includegraphics[width=8.5cm]{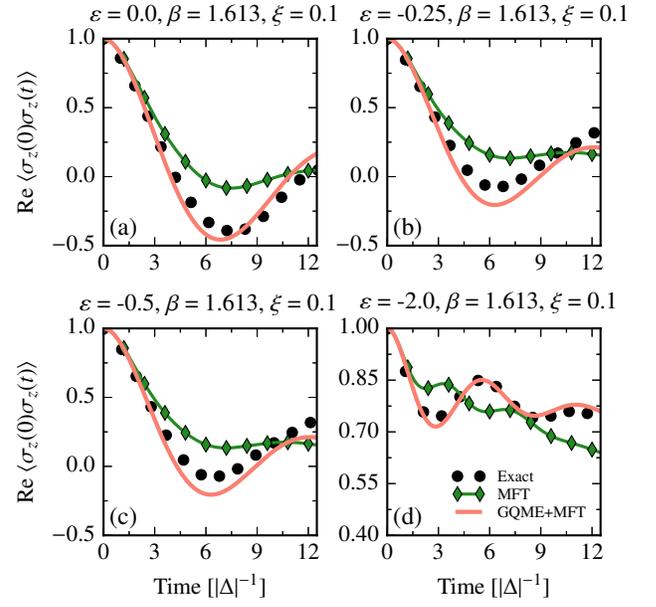} 
\centering
\caption{Comparison of $\mathcal{C}_{zz}(t) = \mathrm{Re}\ \langle \sigma_z(0)\sigma_z(t)\rangle$ obtained using the $cf0$ memory kernels for the biased SB model with $\Delta = -1$, $\omega_c = 14.0$, and $\alpha = -1$.  For all panels (a)-(d) $N = 2$ path integral slices were taken to obtain a converged representation of the canonical density. Exact results are obtained from Ref.~\onlinecite{Egger1992}.}\label{Ch4Fig:Czz_EW}
\end{figure}

Before continuing to the ECFs, we note the almost Markovian appearance of the memory kernels shown in Fig.~\ref{Ch4Fig:K_cf0}.  Panels (a) and (b) correspond to an unbiased ($\varepsilon = 0$) system coupled to a moderately fast bath ($\omega_c = 2.5$), and panels (c) and (d) to a strongly biased system ($\varepsilon = -2$) coupled to a fast bath ($\omega_c =15$).  In both cases shown, the diagonal components of the memory kernel $\mathcal{K}_{xx}(t)$ and $\mathcal{K}_{yy}(t)$ are strongly peaked as $t= 0$ and decay quickly to zero (see panels (a) and (c)).  Interestingly, the off-diagonal components $\mathcal{K}_{xy}(t)$ and $\mathcal{K}_{yx}(t)$ behave very differently; panel (b) shows memory kernels that fluctuate around zero with a small amplitude, while panel (d) displays a high-amplitude short-time peak, which indicates that the off-diagonal components of the memory kernel play a large role in the relaxation of the ECF dynamics for systems characterized by a large bias.  Also, consistent with our expectations, the realization of the SB model in panels (a) and (b) characterized by a moderately fast bath ($\omega_c = 2.5$) exhibits a longer-lived memory kernel than the realization in panels (c) and (d) which corresponds to a faster bath ($\omega_c = 15$).  Figs.~\ref{Ch4Fig:Czz_MC}(a) and \ref{Ch4Fig:Czz_EW}(d) present the GQME+MFT dynamics obtained from the memory kernels in panels (a) and (b), and (c) and (d), respectively.


As Figs.~\ref{Ch4Fig:Czz_MC} and \ref{Ch4Fig:Czz_EW} illustrate, the advantages afforded by the GQME formalism for nonequilibrium averages are transferable to equilibrium situations. For example, Fig.~\ref{Ch4Fig:Czz_MC} illustrates that in the weak coupling regime shown in panels (a) and (b), the GQME approach is able to easily reproduce  the already accurate dynamics produced via the Ehrenfest method. In these instances, the GQME+MFT scheme also affords boosts in efficiency, since the memory kernel cutoffs used to recover the appropriate dynamics were $\tau_c = 0.5$ and $\tau_c = 1.0$ for panels (a) and (b), respectively. Naturally, the limited applicability of the Ehrenfest method can also lead to dynamics that significantly deviate from the exact results. Indeed, by virtue of its mean-field character, the Ehrenfest scheme is known to fail for cases where the system-bath coupling is large, as shown in panels (c) and (d).  Here the Ehrenfest scheme leads to overly fast relaxation. In contrast, the GQME+MFT method is able to produce dynamics that are in notable agreement with the numerically exact results, albeit at a similar computational cost as the direct Ehernfest calculation for panels (c) and (d), where $\tau_c = 3.0$. Further, it is remarkable that even for unbiased cases, where the Ehrenfest method already performs well for nonequilibrium dynamics, the GQME+semiclassics approach can offer marked improvements in the equilibrium case.

The approximations that underlie the Ehrenfest approach also lead to incorrect dynamics in other parameter regimes.  For instance, the classical treatment of the bath implies that the Ehrenfest scheme is most accurate for slow baths where quantum effects are negligible \cite{Stock1995}.  In addition, the mean-field character and classical treatment of the bath in the Ehrenfest framework results in the breaking of detailed balance, a problem that becomes most pronounced in the dynamics of biased systems \cite{Parandekar2005, Parandekar2006}. Cases that lie beyond the region of validity of the Ehrenfest method are also those where the improvements afforded by the GQME+MFT approach are most dramatic, as Fig.~\ref{Ch4Fig:Czz_EW} illustrates.  Given the above considerations, it is not surprising that the Ehrenfest method can only qualitatively capture the relaxation of the ECF, leading to overly fast relaxation and, in the case of the strongly biased system in panel (d), leading to the wrong oscillation frequency.  In contrast, the dynamics produced by the GQME+MFT approach are in remarkable agreement with the numerically exact results.  Hence, as  in the nonequilibrium case, the present method represents an important tool that has the potential to correct the semiclassical dynamics of systems characterized by nonzero bias or coupling to fast baths.  


One question continues to emerge from the continued success of the GQME+semiclassics formalism: where do such improvements in accuracy over bare approximate dynamics come from? In the first paper of this series \cite{Montoya2016a} we conjectured that the source was, at least to some extent, the \textit{exact} sampling of \textit{distinct} initial conditions required for the calculation of the auxiliary kernels. As our added emphasis indicates, this improvement appears to rely on two different factors. The latter, in the nonequilibrium case, was identified as the ``correction terms'' arising from the Wigner transform of the product of the bath part of the system-bath interaction with the bath distribution. In the nonequilibrium case, this operator takes the form $\zeta^W_{noneq} = -\alpha \sum_k c_k P_k \tanh(\beta \omega_k/2)/\omega_k$ (see Eq.~(D6) in Appendix D of Ref.~\onlinecite{Montoya2016a}).  Our treatment of the equilibrium problem produces the analogous bath operator given by Eq.~(\ref{Ch4Eq:ZetaEq}) in Appendix \ref{Ch4App:AuxKernels}. The inclusion of the correlation functions that require the sampling of this operator in the auxiliary kernels (see Eqs.~(\ref{Ch4Eq:EH_C_xm})-(\ref{Ch4Eq:S_asym})) is indeed necessary to obtain the improved dynamics shown in Figs.~\ref{Ch4Fig:Czz_MC} and \ref{Ch4Fig:Czz_EW}. The importance of the first factor, i.e., the \textit{exact} nature of the sampling at $t=0$, in the context of thermal equilibrium can only be confirmed by using representations of the canonical density that can be made arbitrarily accurate.  It bears noting that, for realizations of the SB model where the system and bath are weakly coupled and at high temperatures, the canonical distribution can be captured by a simple factorization approximation implemented earlier in the literature \cite{Wang1998b}.  The two simplest approximations consist of rewriting the Boltzmann factor as simple products of system and bath operators, e.g., $e^{-\beta H} \approx e^{-\beta H_S} e^{-\beta H_B}$ and $e^{-\beta H} \approx e^{-\beta (H_B+H_{SB})/2}e^{-\beta H_S} e^{-\beta (H_B+H_{SB})/2}$, which we refer to in our path integral notation as containing $N =0$ and $N=1$ path integral slices, respectively. For cases where the system-bath coupling is strong (large $\xi$) or temperatures are low (large $\beta$), the Boltzmann factor can no longer be captured by such crude approximations.  In these cases, we adopt the path integral scheme developed in Ref.~\onlinecite{Montoya2016c} to obtain highly accurate expressions for the canonical density, $\rho$. 

\begin{figure}
\centering
\includegraphics[width=8.5cm]{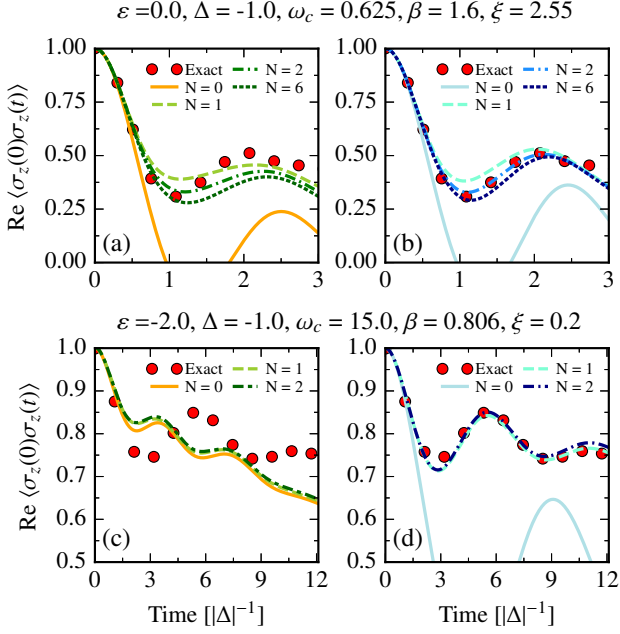} 
\caption{Convergence properties of the direct Ehrenfest and GQME+MFT approaches with respect to the number of path integral slices used in constructing the canonical density operator.  Panels (a) and (c) correspond to direct Ehrenfest dynamics and panels (b) and (d) correspond to the analogous the GQME+MFT results.  Further, for panels (a) and (b) $\alpha = 1$, while for panels (c)-(d) $\alpha = -1$.}\label{Ch4Fig:Ndivs}
\end{figure}

In Fig.~\ref{Ch4Fig:Ndivs}, two sets of dynamics are presented.  Panels (a) and (c) correspond to the direct Ehrenfest treatment of $C_{zz}(t)$ for two different realizations of the SB model using an increasing number, $N$, of path integral slices.  As is evident in panel (a), which corresponds to a case with strong system-bath coupling ($\xi = 2.55$) and moderate to low temperatures ($\beta = 1.6$), a large number of path integral steps ($N = 6$) is required to achieve convergence. Conversely, the high temperature ($\beta = 0.806$) and weak coupling ($\xi = 0.2$) case presented in panel (c) only requires two path integral steps ($N =2$) to achieve convergence.  More importantly, panels (a) and (c) illustrate that the accurate representation of the Boltzmann factor can significantly improve the dynamics produced via the \textit{bare} Ehrenfest method.  Panels (b) and (d) present the GQME+MFT dynamics obtained using the different approximations for $\rho$.  As the figure suggests, the accurate rendering of the canonical distribution is critical in obtaining improved accuracy via the GQME framework.  Moreover, the sensitive dependence of the improvements afforded by the GQME formalism on the accurate rendering of $\rho$ lends credence to our conjecture \cite{Montoya2016a} that the \textit{exact} sampling of \textit{distinct} initial conditions required by the auxiliary kernels is largely responsible for the improvements in accuracy afforded by the GQME framework.

\subsection{Memory Kernel Closures}
\label{Ch4Subsec:Closures}

As was demonstrated in the first paper of this series \cite{Montoya2016a}, use of different closures can strongly influence the quality of the GQME+semiclassics dynamics for nonequilibrium averages. Here we show that a similar dependence also exists in the equilibrium case. In particular we determine that, while the differences between the forward and backward closures are minor, the forward closures outperform their backward counterparts.  In the following analysis of the closures that employ numerical time derivatives ($cf1$, $cf2$, and $cf3$), we provide further numerical support for the analytical proofs in Ref.~\onlinecite{KellyMontoya2016} regarding closures that can be written in terms of the original ECF and its time derivatives.

\begin{figure}
\centering
\includegraphics[width=8.5cm]{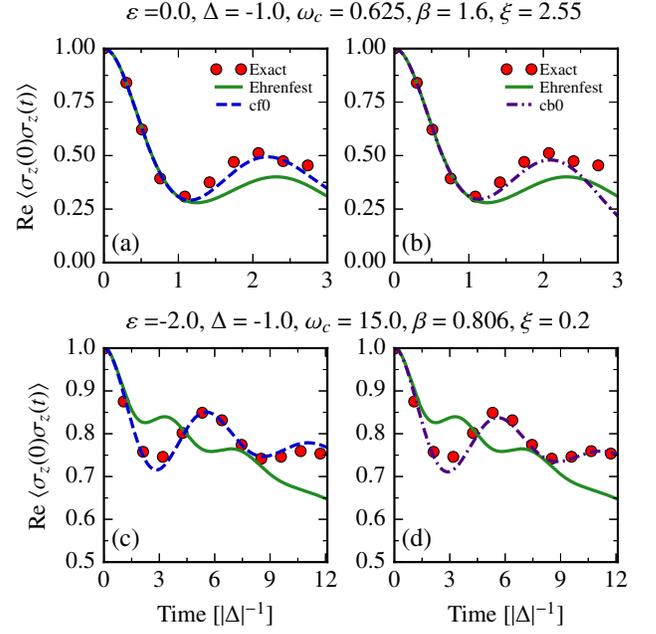} 
\caption{Comparison of $\mathcal{C}_{zz}(t) = \mathrm{Re}\ \langle \sigma_z(0)\sigma_z(t)\rangle$ obtained using the $\mathcal{Q}$-forward  and $\mathcal{Q}$-backward $cf0$ and $cb0$ closures for two realization of the SB model.  For panels (a)-(b), $\alpha = 1$, while for panels (c)-(d) $\alpha = -1$.}\label{Ch4Fig:ForBackCz}
\end{figure}


Initially we focus on the differences between dynamics produced via the $\mathcal{Q}$-forward and $\mathcal{Q}$-backward closures, presented in Fig.~\ref{Ch4Fig:ForBackCz}.  Taking the zeroth order closures $cf0$ and $cb0$, which do not replace the action of the Liouvillian with numerical derivatives, as representative of the differences between the forward and backward closures, it is clear that the forward closure performs slightly better than the backward closure.  This manifests in the difference between the GQME results in panels (a) and (b).  The reason for this discrepancy, however, is not well understood and will require further analysis beyond the scope of the present work.  These differences notwithstanding, both the $cf0$ and $cb0$ closures clearly yield dynamics that are significantly more accurate than the bare quasiclassical results.

\begin{figure}
\centering
\includegraphics[width=8.5cm]{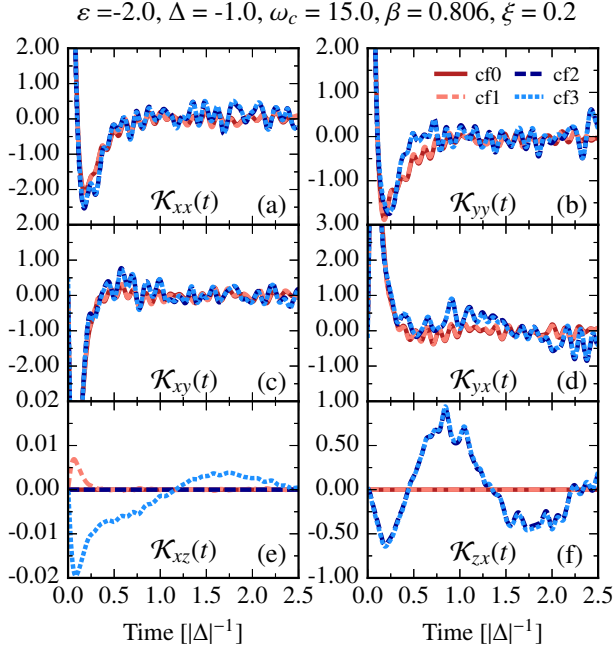} 
\caption{Comparison of the memory kernels obtained using the $\mathcal{Q}$-forward closures $cf0$, $cf1$, $cf2$, and $cf3$ for the SB model with $\alpha = -1$. Note that the $cf1$, $cf2$, and $cf3$ violate the symmetry requiment that stipulates that all memory kernels crossed with $\sigma_z$ must be identically zero.}\label{Ch4Fig:allK}
\end{figure} 
 
Restricting our attention to the forward closures, the kernels produced via the various closures can shed light on the factors that lead to distinct GQME dynamics. As panels (a)-(d) of Fig.~\ref{Ch4Fig:allK} suggest, the closures can be subdivided into two main sets which yield similar memory kernels: the first group consists of closures $cf0$ and $cf1$, while the second of $cf2$ and $cf3$.  The main difference between the kernels produced by these two groups is clearest at intermediate times.  Before individually considering the kernel elements in panels (e) and (f),  it is noteworthy that all components of the memory kernel that depend on $\sigma_z$, e.g., $\mathcal{K}_{xz}(t)$ and $\mathcal{K}_{zx}(t)$, must be zero by symmetry requirements.\footnote{This becomes clear upon considering that $\mathcal{K}(t) = \mathcal{Y}^T (\delta V_B \mathbf{\sigma}|e^{i\mathcal{Q}\mathcal{L}t}| \delta V_B \mathbf{\sigma})\mathcal{Y}$ and that the components of the static transformation matrices $\mathcal{Y}^{T}_{zi} = 0 = \mathcal{Y}^{T}_{iz}$, where $i \in \{x,y,z \}$.}  In light of this, it is apparent that the $cf1$ closure leads to a violation of this symmetry requirement for $\mathcal{K}_{xz}(t)$ (and $\mathcal{K}_{xz}(t)$, not shown) in panel (e), while the $cf2$ closure leads to a similar violation for $\mathcal{K}_{zx}(t)$ (and $\mathcal{K}_{zy}(t)$, not shown) in panel (f).  As the amplitudes of the memory kernels in panels (e) and (f) indicate, this violation is milder for $cf1$ than for $cf2$.  The worst of these closures seems to be $cf3$, for which all $z$-components of the memory kernel are nonzero.  To understand the source of these violations, it is sufficient to consider that while the symmetry requirements are satisfied via the analytical application of $\mathcal{Q}\mathcal{L}|\mathbf{\sigma})$ and $(\mathbf{\sigma}|\mathcal{L}\mathcal{Q}$, replacement of the Liouvillian with the numerical time derivative does not ensure this strict requirement.  On the other hand, the severity of this artificial nonzero behavior cannot be estimated \textit{a priori}.

\begin{figure}
\centering
\includegraphics[width=8.5cm]{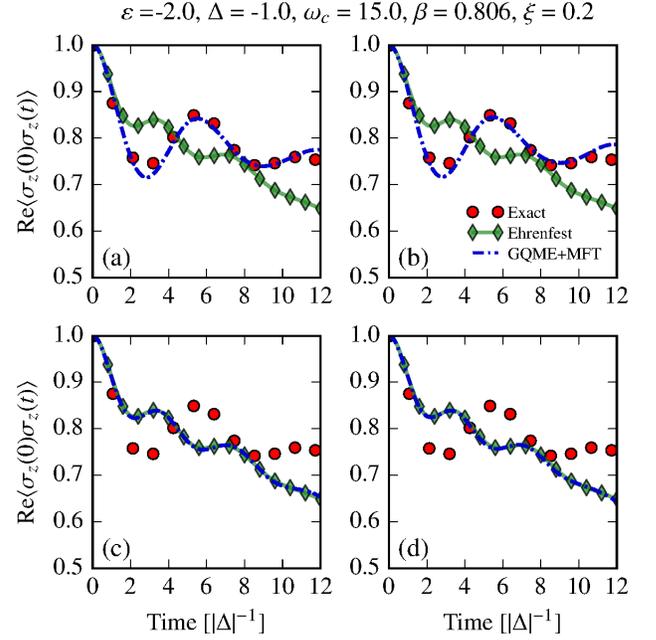} 
\caption{Comparison of $\mathcal{C}_{zz}(t) = \mathrm{Re}\ \langle \sigma_z(0)\sigma_z(t)\rangle$ obtained using the $\mathcal{Q}$-forward  and $\mathcal{Q}$-backward $cf0$ and $cb0$ closures for two realization of the SB model.  For panels (a)-(b), $\alpha = 1$, while for panels (c)-(d) $\alpha = -1$.}\label{Ch4Fig:CzAllClosures}
\end{figure}


The consequences of the differences among the different kernels can be best appreciated in the dynamics they produce, which are shown in Fig.~\ref{Ch4Fig:CzAllClosures}.  As panels (a) and (b) suggest, the $cf0$ and $cf1$ closures lead to the dynamics that most closely reproduce the exact results.  Further, the exceptional agreement between the dynamics produced via the $cf0$ and $cf1$ closures indicates that the slight symmetry violation in Fig.~\ref{Ch4Fig:allK}(c) does not introduce serious errors in the GQME results.  A similar conclusion can be drawn from the dynamics produced by closures $cf2$ and $cf3$, which are able to reproduce the Ehrenfest results to within graphical accuracy.  Importantly, this agreement provides additional numerical support for the analytical arguments put forth in Ref.~\onlinecite{KellyMontoya2016}.  Specifically, because the auxiliary kernels for closure $cf3$ (and $cb3$) can be written in terms of the original ECF in Eq.~(\ref{Ch4Eq:GQME}) and its time derivatives, the memory kernel obtained using this closure \textit{must} recover the bare Ehrenfest result. Because the $cf2$ (and $cb2$) closure also reproduces the Ehrenfest dynamics, as was the case in the nonequilibrium problem \cite{Montoya2016a}, it is clear that the Ehrenfest method permits replacing the action of the Liouvillian operator acting on a dynamically evolved operator with its time derivative.  In turn, the fact that the $cf0$ and $cf1$ (and $cb0$ and $cb1$) closures yield GQME dynamics that are distinct from the Ehrenfest results is consistent with the observation that the Ehrenfest procedure is not ensemble conserving. Nevertheless, the arguments in Ref.~\onlinecite{KellyMontoya2016} do not explain the improvements afforded by the $cf0$ and $cf1$ closures.  Instead, we again emphasize that the most likely reason for the improvement afforded by the $cf0$ and $cf1$ closures lies in the \textit{exact} sampling of \textit{distinct} bath operators alluded to in Sec.~\ref{Ch4Subsec:EffieciencyAccuracy}.

\subsection{Long-Time Dynamics}
\label{Ch4Subsec:LongTimeDynamics}

\begin{figure}
\centering
\includegraphics[width=8.5cm]{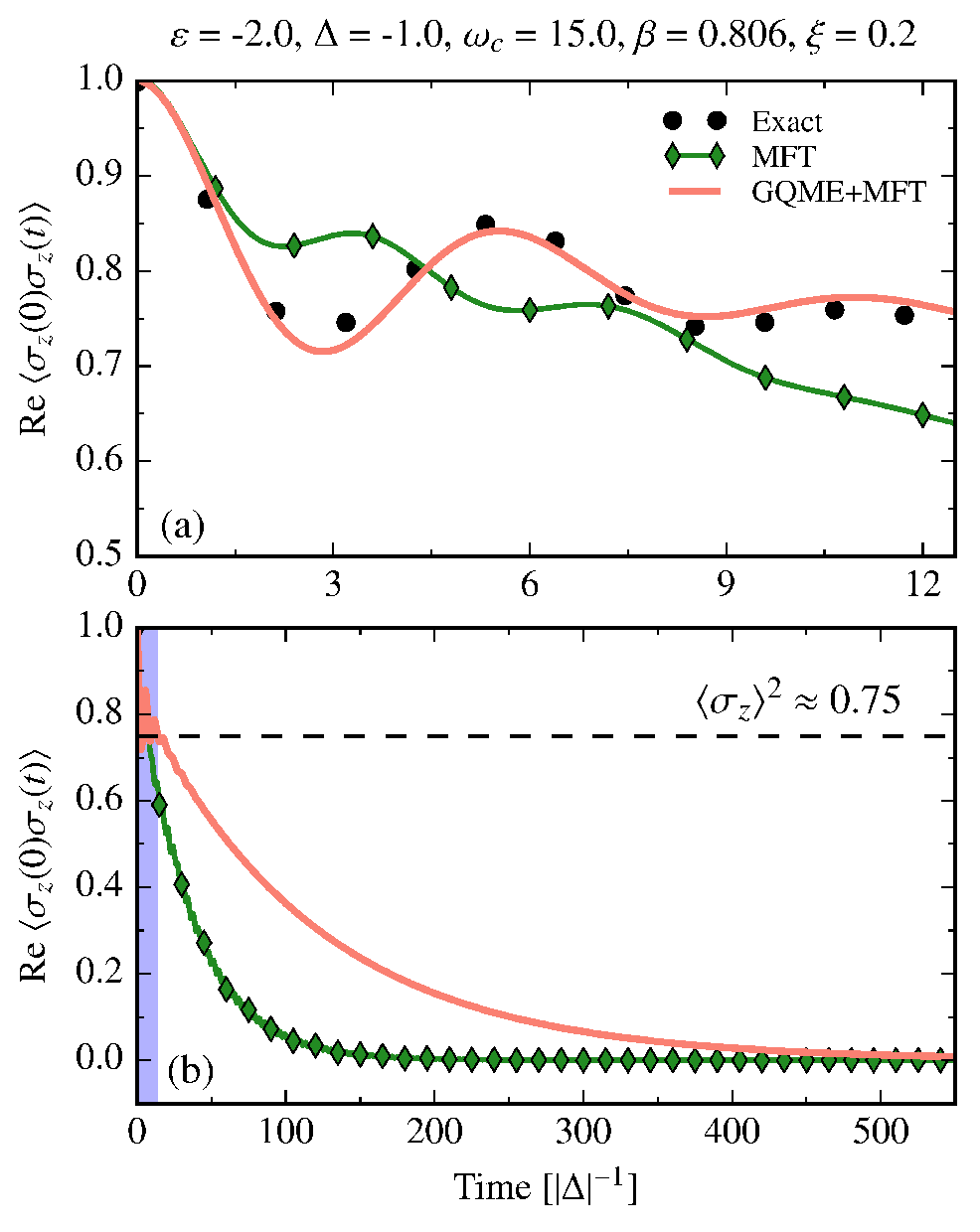} 
\caption{Short- and long-time dynamics for $\mathcal{C}_{zz}(t) = \mathrm{Re}\ \langle \sigma_z(0)\sigma_z(t)\rangle$ obtained using the $cf0$ closure.  Panel (a) illustrates the short-time agreement between the GQME+MFT and exact results.  Panel (b) presents the long-time dynamics obtained via the GQME+MFT and direct Ehrenfest approaches.  The red vertical line indicates the longest time for which numerically exact dynamics are available.  To obtain convergence for the long-time dynamics a memory cutoff time $\tau_c = 12$ was used. }\label{Ch4Fig:LongTimeDynamics}
\end{figure}

 A particularly appealing feature of the GQME approach is the promise of long-time dynamics for a comparatively small computational expense. As shown in Refs.~\onlinecite{Kelly2015} and \onlinecite{Montoya2016a}, the GQME+MFT approach has proven capable of correctly recovering the long-time limit for nonequilibrium averages. In the equilibrium case, however, a different pattern emerges. Recapitulating the results in Fig.~\ref{Ch4Fig:CzAllClosures}(d), panel (a) in Fig.~\ref{Ch4Fig:LongTimeDynamics} illustrates the ability of the GQME+MFT approach to include corrections to the bare Ehrenfest dynamics for intermediate times.  Panel (b), addresses the long-time dynamics for the same set of parameters and contains several important pieces of information.  First, the violet region demarcates the range of time where exact dynamics are available, and the dashed line marks the correct long-time limit of the correlation function, $\mathrm{Re}\langle \sigma_z(t \rightarrow \infty) \sigma_z(0)\rangle = \langle \sigma_z\rangle^2 \approx 0.75$.  This panel demonstrates that the closures that had led to improvements in accuracy for intermediate-time equilibrium dynamics and nonequilibrium averages, i.e., closures $cb0$. $cb1$, $cf0$, and $cf1$, are unable to recover the correct long-time limit for the ECFs. It is also worth noting that the memory cutoff time, $\tau_c$, necessary to recover the long-time dynamics is not necessarily the same as that used before for the intermediate-time dynamics.  Of course, extending $\tau_c$ should not and, indeed, does not appreciably change the intermediate-time dynamics. In this regard, $\tau_c$ is dependent on the choice of time-scales one intends to probe via the GQME. 
 
The reason for the violation of detailed balance at long times by closures $cf0$ and $cf1$ (and $cb0$ and $cb1$) is currently unclear and requires further analysis. Indeed, it may be that the intermediate-time corrections of the memory kernel afforded by the \textit{exact} sampling of \textit{distinct} system-bath operators at $t=0$ are not sufficient to ensure that the long-time dynamics are properly captured.  In line with this argument, there is the possibility of a low-amplitude and long-lasting component of the memory kernel that renders the idea of a well-defined cutoff time for the memory kernel invalid.  Such a component may arise from the fact that the projector used here does not span the whole subspace of the impurity in the SB model, a necessary but not sufficient condition for ensuring that the memory kernel will be short lived (cf.\ the discussion on the choice of Redfield- versus NIBA-type projectors in the nonequilibrium case explored in Ref.~\onlinecite{Montoya2016a}) If this is the case, then the inability of the GQME+MFT approach to obtain the correct long-time limit of the ECF is not a fundamental shortcoming of the method, but rather an unfortunate artifact due to the choice of projection operator. Note that in the case of the nonequilibrium population of Ref.~\onlinecite{Montoya2016a} the projector does span the complete space of the impurity. Because such an analysis goes beyond the scope of the present study, we postpone work in this direction to future papers. It is important to note, however, that the GQME+MFT approach consistently leads to improved intermediate-time dynamics and long-time dynamics that are as good or better than those obtained using the Ehrenfest method directly.  In particular, for unbiased cases the improvements obtained are compatible even for long times with those found in the nonequilibrium case of Ref.~\onlinecite{Montoya2016a}.   Thus, for example, the autocorrelation function of the dipole moment, which lies at the heart of line shape theory \cite{Anderson1954,Kubo1969}, and correlation functions that probe fluctuations away from equilibrium, such as those used to calculate transport coefficients in the linear response regime \cite{MahanBook} would be unaffected by this issue. 

A final point of interest concerns the use of the approach presented here to extend the temporal range of standard semiclassical simulations.  Due to the fact that the statistical error in quasi- and semi-classical methods increases with simulation time, it becomes prohibitively expensive to converge quasi- and semi-classical dynamics on the time-scales studied in Fig.~\ref{Ch4Fig:LongTimeDynamics}(b), i.e., $500 |\Delta|^{-1}$.  To ameliorate this problem, we have instead used the GQME+MFT dynamics obtained using the $cf3$ closure, which is known to reproduce the direct Ehrenfest result.  In this way, it was possible to extend the Ehrenfest dynamics to nontrivially long times for the expense of a simple $5|\Delta|^{-1}$-long Ehrenfest calculation.

\section{Conclusions}
\label{Ch4Sec:Conclusions}

In this paper we have extended the GQME+MFT method introduced in the first paper of this series \cite{Montoya2016a} to treat ECFs within the context of the Mori formalism. As Ref.~\onlinecite{Montoya2016a} and the present work demonstrate, the Mori-based formulation furthered here provides a flexible framework to accurately study problems both in and out of equilibrium for SB-type models.  We further emphasize that this approach is general and can be easily generalized to arbitrarily complex systems and dynamical quantities. 

In the same spirit as earlier work on nonequilibrium dynamics \cite{Shi2003,Zhang2006a,Cohen2011,Cohen2013,Cohen2013a,Wilner2013, Wilner2014,Shi2004a,Kelly2013,Kelly2015,Pfalzgraff2015,Montoya2016a,KellyMontoya2016}, we have demonstrated that equilibrium memory kernels for the SB model can be short-lived in comparison to the ECF lifetime.  Indeed, it is possible to say that \textit{all} impurity-type systems where the bath consists of a large number of modes with a broad distribution of energies have short-lived memory kernels, as long as the projection operator includes all system states.  We have also shown that, as in the nonequilibrium case, the GQME+semiclassics approach is capable of yielding impressive improvements in computational efficiency and accuracy over direct application of bare quasi- and semi-classical methods. Consequently, at least for these types of models, we foresee that the GQME approach can become an invaluable tool for obtaining highly accurate nonequilibrium \textit{and} equilibrium dynamics at a lower computational cost than would be necessary with conventional methods that go well beyond SB-type models, including the dynamics of liquid state systems with no system-bath distinction.

Via a systematic analysis of the dependence of the GQME dynamics on the kernel closures, we have further confirmed the analytical arguments posited in Ref.~\onlinecite{KellyMontoya2016}.  Specifically, the results obtained from the $cf3$ and $cb3$ closures, which reproduce the direct Ehrenfest dynamics, confirm the proof presented in Ref.~\onlinecite{KellyMontoya2016} which states that when the auxiliary kernels can be written in terms of the original ECF and its time derivatives, the GQME results will be equivalent to those arising from direct application of the dynamical method. The fact that the $cf2$ and $cb2$ closures also reproduce the Ehrenfest results is in agreement with the the observation that the numerical time derivative of a correlation function is equivalent to the action of the Liouvillian on the dynamically sampled operator in the Ehrenfest method.  We have also demonstrated that the $cf0$ closure yields the most accurate dynamics over a wide region of parameter space.  Further, although the $cb0$ closure has been shown to produce slightly less accurate dynamics for some sets of parameters, it also affords a significant improvement over the bare mean-field dynamics, as do the $cf1$ and $cb1$ closures, despite the slight violation of the symmetry requirement that requires the memory kernels involving $\sigma_z$ to be zero. The distinct results obtained from the $cf0$, $cb0$, $cf1$, and $cb1$ closures has also been shown to be consistent with the fact that the Ehrenfest method is \textit{not} ensemble-conserving \cite{KellyMontoya2016}.

To conclude, we emphasize that the Mori formulation presented in this work and Ref.~\onlinecite{Montoya2016a} is equally applicable to impurity-type models like the SB model and systems such as spin and fermion lattice models \cite{Rasetti1991, Sachdev2011, Giamarchi2004}, and quantum fluids \cite{Feenberg1969, PinesNozieres1999, Poulsen2005, Markland2011}.  In future papers, we explore the extension of the Mori framework to multiple-time correlation functions and systems coupled to arbitrary baths as well as problems where the system-bath distinction is absent. 

\section{Acknowledgements}
\label{Ch4Sec:Acknowledgements}

The authors would like to thank Aaron Kelly and Tom Markland for helpful discussions. D.R.R. acknowledges support from the NSF Grant No. CHE-1464802.  A.M.C. thanks Hsing-Ta Chen for useful conversations.

\appendix

\section{Path integral treatment of the Wigner transformed canonical density operator}			
\label{Ch4App:PIrho}

After a partial Wigner transform with respect to the bath degrees of freedom, it is possible to write the system components of the canonical density operator as follows \cite{Montoya2016c},
	\begin{equation}\label{Ch4Eq:SchematicRho}
	\begin{split}
	\rho^W_{a,b} &= [2\pi]^{-f}\int d\mathbf{s}\ e^{-i\mathbf{p}\cdot \mathbf{s}/\hbar} \bra{\mathbf{x} + \mathbf{s}/2} \bra{a}\rho\ket{b} \ket{\mathbf{x} - \mathbf{s}/2}\\
	&\equiv N_{ab}\cdot \mathcal{R}_{a,b}^W(\mathbf{x}, \mathbf{p}), 
	\end{split}
	\end{equation}
where $\rho = e^{-\beta H}/\mathrm{Tr}[e^{-\beta H}]$, $a, b \in \{0,1\}$ correspond to the two system states of the SB model, $N_{ab}$ is temperature dependent normalization constant, and $\mathcal{R}_{a,b}^W(\mathbf{x}, \mathbf{p})$ is a bath operator of unit partial trace, i.e., $\int d\mathbf{x} d\mathbf{p}\ \mathcal{R}_{a,b}^W(\mathbf{x}, \mathbf{p}) = 1$, which can be interpreted as the bath distribution function.  We henceforth drop the dependence of the bath distribution function on the bath coordinates and momenta, $(\mathbf{x}, \mathbf{p})$, for notational clarity.

To obtain expressions for $N_{ab}$ and $\mathcal{R}_{a,b}^W$ we follow the numerically exact framework presented in Ref.~\onlinecite{Montoya2016c}, which relies on the path integral expansion of the Boltzmann factor.  This expansion employs the Trotter factorization, allowing us to rewrite the full Boltzmann factor as an $N$-membered product of basic path integral units, 
	\begin{equation}\label{Ch4Eq:Trotter}
	e^{-\beta H} \approx [e^{-\beta (H_B + H_{SB})/2N}e^{-\beta H_{S}/N}e^{-\beta (H_B + H_{SB})/2N}]^N.
	\end{equation}
Rigorously, this approximation becomes exact in the limit of $N \rightarrow \infty$, but in practical calculations convergence is achieved with a small $N$. 

By substituting Eq.~(\ref{Ch4Eq:Trotter}) into Eq.~(\ref{Ch4Eq:SchematicRho}), introducing resolutions of the identity $\mathbf{1}_S = \sum_{a} \ket{a}\bra{a}$ and $\mathbf{1}_B = \int d\mathbf{q} \ket{\mathbf{q}}\bra{\mathbf{q}}$ for the system and bath subspaces, respectively, and integrating over the bath degrees of freedom one can obtain expressions for $N_{ab}$ and $\mathcal{R}_{a,b}^W$. While the previous procedure formally eliminates the dependence on the bath degrees of freedom introduced with the resolutions of the identity, the same cannot be said of the system degrees of freedom.  In fact, it is worthwhile to note that the number of paths grows exponentially with the number of path integral steps, and that each path can be associated with a sequence of indices that in indicate the electronic states visited along said path.  To illustrate this point, consider the simpler case of applying the path integral formalism to the Boltzmann factor corresponding to the isolated two-level system: 
	\begin{equation}
	\begin{split}
	\bra{k_N} e^{-\beta H_S} \ket{k_0} &\approx \sum_{k_1, ..., k_{N-1}}\bra{k_N} e^{-\beta H_S/N} \ket{k_{N-1}} ...\\
	&\qquad  \bra{k_2} e^{-\beta H_S/N} \ket{k_1}\bra{k_1} e^{-\beta H_S/N} \ket{k_0},
	\end{split}
	\end{equation}
	where $k_n \in \{0,1 \}$ corresponds to the state used in the $n^{th}$ path integral step of the expansion.  For this decomposition, one may consider the path to consist of a string of numbers describing the identity of the states taken in the sequence, e.g., in an $N = 3$ expansion with the endpoints fixed, there are $4( = 2^{N-1})$ such sequences, $\{k_3,0,0,k_0\}$, $\{k_3,0,1,k_0\}$, $\{k_3,1,0,k_0\}$, and $\{k_3,1,1,k_0\}$.  In the following, we refer to the sum over the set $\{k_1, ..., k_{N-1} \}$ as the sum over paths.  
	
Before providing expressions for $N_{ab}$ and $\mathcal{R}_{a,b}^W$, we provide a few definitions that render the notation simpler. We begin with the following basic definitions, 
	\begin{align}
	b^{(l)}_{n} &= - \bra{k_n}\sigma_z \ket{k_n} \alpha c_l/\omega_l^2\nonumber \\  
	&=(-1)^{k_n}\alpha c_l/\omega_l^2,\\
	\theta_l &= \frac{\beta \hbar \omega_l}{2N},\\
	\delta b^{(l)}_{nm} &= b^{(l)}_{n} - b^{(l)}_{m}.
	\end{align}
and the following path-independent quantities, 
\begin{align}
	\gamma^{(l)}_{p}  &= \frac{\tanh(2\theta_l)}{\omega_l \eta_l}, \\
	\gamma^{(l)}_{x} &= \frac{\omega_l \nu_l }{\tanh(2\theta_l)},
	\end{align} 
	\begin{align}
	\eta_l &= 1-\frac{[\mathbf{A}_{l}^{-1}]_{1,1} - [\mathbf{A}^{-1}_l]_{1,N-1}}{\cosh^2(2\theta_l)},\\
	\nu_l &= 1-\frac{[\mathbf{A}_{l}^{-1}]_{1,1} + [\mathbf{A}_l^{-1}]_{1,N-1}}{\cosh^2(2\theta_l)},
	\end{align}
where $\mathbf{A}^{(l)}$ is a tridiagonal $N-1 \times N-1$ matrix whose diagonal and off-diagonal entries are equal to $2$ and $-\mathrm{sech}(2\theta_l)$, respectively.

A few quantities depend on the path taken in configuration space.  These quantities, labeled by a tilde, take the following forms, 
	\begin{align}
	\tilde{\mathcal{S}}_{k_N, k_0} &= \prod_{j = 1}^{N} \bra{k_j}e^{-\beta H_{ad}/N} \ket{k_{j-1}}, \label{Ch4Eq:SPI}
	\end{align}
	\vskip5pt
	\begin{equation}
	\tilde{\boldsymbol{\delta}}^{(l)} = \frac{\mathbf{j}_{l}^T \cdot \mathbf{A}_l^{-1}}{\cosh(\theta_l)}, 
	\end{equation}
	\begin{equation}
	\begin{split}
	\tilde{\mathbf{j}}_l &= \left[ \begin{array}{c}
\Delta b^{(l)}_{21} - \Delta b^{(l)}_{10}   \\
\Delta b^{(l)}_{32} - \Delta b^{(l)}_{21}  \\
\vdots \\
\Delta b^{(l)}_{N,N-1} - \Delta b^{(l)}_{N-1,N-2}   \end{array}\right],   
\end{split}
	\end{equation}

\begin{widetext}
	\begin{equation}
	\begin{split}
	\tilde{\kappa}_{p}^{(l)} =  \left\{
      \begin{array}{lr}
       -\frac{\omega_l}{2\tanh(2\theta_l)}\Bigg[\frac{\cosh(\theta_l)}{\cosh(2\theta_l)}[(\Delta b^{(l)}_{N,N-1} +\Delta b^{(l)}_{1,0})-(\tilde{\boldsymbol{\delta}}^{(l)}_{N-1}-\tilde{\boldsymbol{\delta}}^{(l)}_1)] - \eta_l \Delta b^{(l)}_{N,0} \Bigg] \quad &: \quad N \geq 2,\\
       -\frac{\omega_l}{\tanh(2\theta_l)}\Bigg[\frac{\cosh(\theta_l)}{\cosh(2\theta_l)}\Bigg]\Delta b^{(l)}_{1,0} [1 - \cosh(\theta_l)]  &: \quad N=1,
      \end{array}
    \right.
	\end{split}
	\end{equation}
	\begin{equation}
	\begin{split}
	\tilde{\kappa}_{x}^{(l)} =  \left\{
      \begin{array}{lr}
       \frac{\cosh(\theta_l)}{2\cosh(2\theta_l)}\frac{  [(\Delta b^{(l)}_{N,N-1} - \Delta b^{(l)}_{1,0}) - (\tilde{\boldsymbol{\delta}}^{(l)}_{N-1}+\tilde{\boldsymbol{\delta}}^{(l)}_1)]}{\nu_l} - \frac{b^{(l)}_N + b^{(l)}_0}{2} \quad &: \quad N \geq 2,\\
       - \frac{b^{(l)}_N + b^{(l)}_0}{2}  &: \quad N=1,
      \end{array}
    \right. 
	\end{split}
	\end{equation}
	\begin{equation}
	\begin{split}
	\tilde{\Lambda}^{(l)} =  \left\{
      \begin{array}{lr}
       \frac{\omega_l}{4 \tanh(2\theta_l)} \Bigg[  \frac{1 + \cosh(2\theta_l)}{\cosh(2\theta_l)}\Big[\sum_{j = 1}^{N} [\tilde{\boldsymbol{\delta}} b^{(l)}_{j,j-1}]^2 -\frac{\tilde{\mathbf{j}}_l^T \cdot \mathbf{A}_l^{-1} \cdot \tilde{\mathbf{j}}_l}{\cosh(2\theta_l)}\Big]   - 2 \frac{\cosh(\theta_l)}{\cosh(2\theta_l)}\Big[(\Delta b^{(l)}_{N,N-1}+\Delta b^{(l)}_{1,0})-(\tilde{\boldsymbol{\delta}}^{(l)}_{N-1}-\tilde{\boldsymbol{\delta}}^{(l)}_1)\Big]\Delta b^{(l)}_{N,0} \quad &  \\
       \qquad \qquad  \qquad - \Big[\frac{\cosh(\theta_l)}{\cosh(2\theta_l)}\Big]^2\frac{ [(\Delta b^{(l)}_{N,N-1} - \Delta b^{(l)}_{1,0}) - (\tilde{\boldsymbol{\delta}}^{(l)}_{N-1}+\tilde{\boldsymbol{\delta}}^{(l)}_1)]^2}{\nu_l} + \eta_l [\Delta b^{(l)}_{N,0}]^2\Bigg] \quad &: \quad N \geq 2,\\
       \frac{\omega_l}{\tanh(2\theta_l)}\Bigg[\frac{\cosh(\theta_l)}{\cosh(2\theta_l)}\Bigg][\Delta b^{(l)}_{1,0}]^2 [1 - \cosh(\theta_l)]  &: \quad N=1,
       \end{array}
    \right. 
	\end{split}
	\end{equation}

Using these definitions,  
	\begin{equation}\label{Ch4Eq:WignerR}
	\begin{split}
	\mathcal{R}_{a,b}^W &= \Bigg[ \prod_{l = 1}^{F} \frac{\sqrt{\nu_l/\eta_l}}{\pi }\Bigg] \sum_{\mathrm{paths}} \frac{\tilde{W}_{a,b}}{\mathcal{W}_{a,b}}    \prod_{l = 1}^{F} \exp\Big[-\gamma_p^{(l)}( p_l + i\tilde{\kappa}_p^{(l)})^2 -\gamma_x^{(l)} (x_l + \tilde{\kappa}_x^{(l)})^2 \Big],
	\end{split}
	\end{equation}
	\begin{equation}
	\begin{split}
	N_{ab} &=\frac{\mathcal{W}_{a,b}}{\sum_a  \mathcal{W}_{a,a}},
	\end{split}
	\end{equation}
	where $\tilde{W}_{a...b} = \tilde{\mathcal{S}}_{a,b} \exp[-\sum_l \tilde{\Lambda}^{(l)}]$, and $\mathcal{W}_{a,b} = \sum_{paths} \tilde{W}_{a,b}$.

Finally, in the following section, the Wigner transform of the product of $\rho$ and $V_B = \alpha \sum_l c_l x_l$ will require the evaluation of the following operator, 
	\begin{equation}\label{Ch4Eq:ZetaEq}
	\begin{split}
	\zeta &= \frac{1}{2 \mathcal{R}_{a,b}^{W} } \frac{\partial \mathcal{R}_{a,b}^{W}}{\partial \mathbf{p}} \cdot \frac{V_B}{\partial \mathbf{x}}\\
	&= -\alpha [\mathcal{R}_{a,b}^W]^{-1}\Bigg[ \prod_{l = 1}^{F} \frac{\sqrt{\nu_l/\eta_l}}{\pi }\Bigg]  \sum_{\mathrm{paths}} \frac{\tilde{W}_{a,b}}{\mathcal{W}_{a,b}} \Big[ \sum_qc_q \gamma_p^{(q)} (p_q + i \tilde{\kappa}^{(q)}_p)\Big]   \prod_{l = 1}^{F} \exp\Big[-\gamma_p^{(l)}( p_l + i\tilde{\kappa}_p^{(l)})^2 -\gamma_x^{(l)} (x_l + \tilde{\kappa}_x^{(l)})^2 \Big].
	\end{split}
	\end{equation}
\end{widetext}

\section{Expressions for the Auxiliary kernels}
\label{Ch4App:AuxKernels}

We first introduce a notation based on static and dynamic matrices that facilitate the construction of the auxiliary kernels. Given the definition of the inner product in Eq.~(\ref{Ch4Eq:InnerProduct}), one can easily verify that 
	\begin{align}
	(i\mathcal{Q}\mathcal{L})|\boldsymbol{\sigma}) &= |\delta V_B \boldsymbol{\sigma})\mathcal{Y}, \\
	 (\boldsymbol{\sigma}|(-i\mathcal{L}\mathcal{Q}) &= \mathcal{Y}^T(\delta V_B \boldsymbol{\sigma}|\label{Ch4Eq:RotatedSigma}
	\end{align}
where $\mathcal{Y}_{nm} = -2\epsilon_{znm}$ is a static transformation matrix, $\epsilon_{ijk}$ is the Levi-Civita tensor, the $T$ superscript in Eq.~(\ref{Ch4Eq:RotatedSigma}) denotes the transpose operation, and $\delta V_B = V_B - \langle V_B \rangle$ is the fluctuation of the interaction with the the bath from its equilibrium value.  It is also necessary to evaluate one more static matrix for the GQME evolution in Eq.~(\ref{Ch4Eq:GQME}),
	\begin{align}
	\dot{\mathcal{C}}_{nm}(0) &= -2\Big[(\varepsilon + \langle V_B \rangle )\epsilon_{znm} + \Delta \epsilon_{xnm} \Big].
	\end{align}
The dynamic matrices are  
	\begin{equation}\label{Ch4Eq:DynamicMatrices}
	\begin{split}
	&\mathcal{C}^{(jk)}_{nm}(t) \equiv (V_{B,j}\sigma_n| V_{B,k}(t)\sigma_m(t)), 
	\end{split}
	\end{equation}
where 
	\begin{equation}
	V_{B,k} = \left\{ \begin{array}{cc}
	V_B , \qquad &k = 1,\\
	1, \qquad &k =0. \end{array} \right.
	\end{equation}
	We further note that $\mathcal{C}^{(00)}_{nm}(t) = \mathcal{C}_{nm}(t)$.  
	
Using these expressions, one can rewrite the auxiliary kernels thus, 
	\begin{align}
	\mathcal{K}^{(1)}(t) &= \mathcal{Y}^T[ \mathcal{C}^{(11)}(t) + \langle V_B \rangle^2 \mathcal{C}^{(00)}(t) \nonumber  \\
	&\qquad \qquad - \langle V_B \rangle (\mathcal{C}^{(01)}(t) + \mathcal{C}^{(10)}(t)) ]\mathcal{Y},\\
	\mathcal{K}^{(3b)}(t) &= \mathcal{Y}^T [\mathcal{C}^{(10)}(t) - \langle V_B \rangle \mathcal{C}^{(00)}(t)],\\
	\mathcal{K}^{(3f)}(t) &= - [\mathcal{C}^{(01)}(t) - \langle V_B \rangle \mathcal{C}^{(00)}(t)]\mathcal{Y}.
	\end{align}
	 
\subsection{Quasi-classical treatment of $\mathcal{C}^{(jk)}_{nm}(t)$}
\label{Ch4App:QuasiClassicalC}
	
	In analogy to the definition of nonequilibrium averages \cite{Grunwald2009}, one may express a correlation function within the Ehrenfest formalism as 
	\begin{equation}
	\begin{split}
	C_{XY}(t) &= \mathrm{Tr}[X_S(0)X_B(0)Y_S(t)Y_B(t)]\\
	&\approx [2\pi \hbar]^{-N} \int d\Gamma\ X_B^W(0) Y_B^W(t)\mathrm{Tr}_S[X_S(0)Y_S(t)],   
	\end{split}
	\end{equation}	 
	where the superscript $W$ indicates the partial Wigner transform of the operator with respect to the bath degrees of freedom, and $N$ is the number of degrees of freedom over which the Wigner transform is being performed. The Wigner transform of an operator is defined as \cite{Hillery1984a},
	\begin{equation}
	X^W(\mathbf{x}, \mathbf{p}) = \int d\mathbf{s}\ e^{-i\mathbf{p}\cdot \mathbf{s}/\hbar} \bra{\mathbf{x} + \mathbf{s}/2} \hat{X} \ket{\mathbf{x} - \mathbf{s}/2}.
	\end{equation}
	In the following, we only perform the partial Wigner transform with respect to the bath degrees of freedom, as is required by the Ehrenfest procedure.
	
	For equilibrium correlation functions of the form given by Eq.~(\ref{Ch4Eq:DynamicMatrices}), it is necessary to obtain an expression for the partially transformed canonical density operator $\rho = e^{-\beta H} / \mathrm{Tr}[e^{-\beta H}]$.  Here we adopt the approach presented in Ref.~\onlinecite{Montoya2016c}, and express partial Wigner transform of the canonical operator as
	\begin{equation}
	\rho^W = \sum_{a} N_{a} \mathcal{S}_{a}  \otimes \mathcal{B}_{a}^W(\mathbf{x}, \mathbf{p}), 
	\end{equation}
where $\mathcal{S}_a$ is a pure system operator, and $\mathcal{B}_{a}^W$ is corresponds to a normalized bath density operator, i.e., $\int d\Gamma\ \mathcal{B}_{a}^W(\mathbf{x}, \mathbf{p}) = 1$ (for detailed expressions, see Appendix \ref{Ch4App:PIrho}).  For convenience, we reexpress $\mathcal{S}_{a}$ in terms of a convenient basis, namely $\mathcal{S}_a = \sum_b r_{ab} S_b$, where $S_b \in \{\mathbf{1}_S, \sigma_x, \sigma_y, \sigma_z \}$.  Using the above notation, we rewrite the expressions for the dynamic matrices in Eq.~(\ref{Ch4Eq:DynamicMatrices}) as follows,
	\begin{equation}\label{Ch4Eq:EhCpre0}
	\begin{split}
	\mathcal{C}^{(jk)}_{nm}(t) &\approx \frac{1}{2}\sum_p \int d\Gamma\ \Bigg[ i\Big([\mathcal{B}_p, V_{B,j}]\Big)^W V_{B,k}^W(t) \mathrm{Tr}_S[X_{pn}\sigma_m(t) ]  \\
	& \qquad \qquad +\Big(\{\mathcal{B}_p, V_{B,j} \}\Big)^W V_{B,k}^W(t) \mathrm{Tr}_S[Y_{pn}\sigma_m(t) ] \Bigg],
	\end{split}
	\end{equation}
where
	\begin{align}
	X_{pn} = \sum_{l\in x,y,z} r_{pl} \epsilon_{lns}\sigma_s,\\
	Y_{pn} = r_{pn} \mathbf{1}_S  + r_{p1} \sigma_n 
	\end{align}
	
Noting that the Wigner transform of operator products can be obtained via the Moyal expansion \cite{Hillery1984a}
	\begin{equation}\label{Ch4Eq:WignerProduct}
	(AB)^W = A^W e^{\hbar\Lambda/2i} B^W, 
	\end{equation}
	where 
	\begin{equation}
	\Lambda = \overleftarrow{\nabla}_p \cdot \overrightarrow{\nabla}_x - \overleftarrow{\nabla}_x \cdot \overrightarrow{\nabla}_p,
	\end{equation}
we can express the commutator and anticommutator of the bath density operator and the bath part of the interaction as, 
	\begin{align}\label{Ch4Eq:EhCpre1}
	\Big(\mathcal{B}_p V_{B,j}\Big)^W  &= \mathcal{B}_p^W\Big[V_{B,j}^W-i\zeta_{B,j}^W\Big], 
	\end{align}
	where
	\begin{equation}
	\zeta_{B,j}^W = \left\{ \begin{array}{cc}
	\zeta_B^W , \qquad &j = 1,\\
	0, \qquad &j =0. \end{array} \right.
	\end{equation}
The expression for $\zeta_B^W$ can be found in Eq.~(\ref{Ch4Eq:ZetaEq}).
	
Substituting expressions (\ref{Ch4Eq:EhCpre1}) into (\ref{Ch4Eq:EhCpre0}) yields, 
	\begin{equation}
	\begin{split}
	\mathcal{C}^{(jk)}_{nm}(t) &= \sum_p \int d\Gamma\ \mathcal{B}_p^W V_{B,k}^W(t) \Big[ \zeta_{B,j}^W\mathrm{Tr}_S[X_{pn}\sigma_m(t) ] \\
	&\qquad \qquad \qquad \qquad \  +V_{B,j}^W \mathrm{Tr}_S[Y_{pn}\sigma_m(t) ] \Big],
	\end{split}
	\end{equation}
	
Explicitly evaluating $X_{pn}$ and $Y_{pn}$, we provide explicit expressions for the  $\mathcal{C}_{nm}^{jk}(t)$-functions in terms of functions that are simple to simulate, 
	\begin{equation}\label{Ch4Eq:EH_C_xm}
	\begin{split}
	\mathcal{C}_{xm}^{(ij)} &= \sum_k [r_{k1} s_{xm,s}^{(ij,k)}(t) + r_{kx} s_{1m,s}^{(ij,k)}(t) \\
	&\qquad \qquad + r_{kz} s_{ym,a}^{(ij,k)}(t) - r_{ky} s_{zm,a}^{(ij,k)}(t)],
	\end{split} 
	\end{equation}
	\begin{equation}\label{Ch4Eq:EH_C_ym}
	\begin{split}
	\mathcal{C}_{ym}^{(ij)} &=  \sum_k [r_{k1} s_{ym,s}^{(ij,k)}(t) + r_{ky} s_{1m,s}^{(ij,k)}(t) \\
	&\qquad \qquad + r_{kx} s_{zm,a}^{(ij,k)}(t) - r_{kz} s_{xm,a}^{(ij,k)}(t)],
	\end{split} 
	\end{equation}
	\begin{equation}\label{Ch4Eq:EH_C_zm}
	\begin{split}
	\mathcal{C}_{ym}^{(ij)} &= \sum_k [r_{k1} s_{zm,s}^{(ij,k)}(t) + r_{kz} s_{1m,s}^{(ij,k)}(t) \\
	&\qquad \qquad  + r_{ky} s_{xm,a}^{(ij,k)}(t) - r_{kx} s_{ym,a}^{(ij,k)}(t)],
	\end{split} 
	\end{equation}
where
	\begin{equation}\label{Ch4Eq:S_sym}
	\begin{split}
	s_{nm,s}^{(ij,k)}(t) &= \int d\Gamma\ \mathcal{B}_{B,k}^W  V_{B,i}^W(0) V_{B,j}^W(t) \mathrm{Tr}_S[ B_n(0) \sigma_m(t)],
	\end{split}
	\end{equation}
	\begin{equation}\label{Ch4Eq:S_asym}
	\begin{split}
	s_{nm,a}^{(ij,k)}(t) &= \int d\Gamma\ \mathcal{B}_{B,k}^W \xi_{B,i}^W(0) V_{B,j}^W(t) \mathrm{Tr}_S[ B_n(0) \sigma_m(t)],
	\end{split}
	\end{equation}
are real and $B_n \in \{ \mathbf{1}_S, \sigma_x, \sigma_y, \sigma_z\}$.  Direct evaluation of these functions is straightforward via the Ehrenfest method.  We direct the reader to Ref.~\onlinecite{Montoya2016a} for instructions on how to properly treat the system initial conditions for these functions.

\subsection{Ehrenfest method}
\label{App:Ehrenfest}

In the Ehrenfest method \cite{Gerber1982, Stock1995, Tully1998a, Grunwald2009}, the system (bath) evolves in the mean field of the bath (system).  Additionally, the dynamics of the bath are assumed to be accurately captured by classical mechanics.  Naturally, the mean field and classical bath approximations imply that the Ehrenfest method is not applicable to cases where the system and bath are strongly coupled (large $\lambda$) and baths characterized by high frequencies (large $\omega_c$) or low temperatures, respectively.  

Classical treatment of the bath requires that the continuous spectral density be discretized into individual oscillators. We use the approach outlined in Ref.~\onlinecite{Craig2005} where the spectral density is decomposed into $f$ oscillators. The frequency of the $k^{th}$ oscillator takes the form, 
	\begin{equation}
	\omega_k = -\omega_c \ln\Bigg[ \frac{k - \frac{1}{2}}{f} \Bigg], 
	\end{equation}
and the coupling constant, 
	\begin{equation}
	c_k = \omega_k  \Bigg[\frac{\xi \omega_c}{f}\Bigg]^{1/2} .
	\end{equation}

Evolution of the system and bath within the Ehrenfest scheme are given by the following equations of motion.  For the system, the Liouville equation under the influence of a modified Hamiltonian determines the evolution of the system's wavefunction, 
	\begin{equation}
	\frac{d}{dt} \rho_S(t) = -i[H_{S}^{Eh}, \rho(t)], 
	\end{equation}
where 
	\begin{equation}
	H_{S}^{Eh}(t) = [\varepsilon + \lambda^{cl}(t)] \sigma_z + \Delta \sigma_x,
	\end{equation}
is the modified system Hamiltonian and $\lambda^{cl}(t) = \alpha \sum_k c_k Q_k(t)$ is the classical fluctuation that modulates the system bias.  Here, $\rho_S(0) = \ket{\psi_S(0)}\bra{\psi_S(0)}$ is the initial density matrix for the system, which must be initialized in a pure state (for a detailed discussion, see Ref.~\onlinecite{Montoya2016a}). 

The evolution of the bath variables is given by Hamilton's equations subject to Hamiltonian of the free bath modified by a quantum force arising from the time-dependent populations in the system, 
	\begin{align}
	\frac{dP_k}{dt}  &= - \frac{\partial H_{B}^{Eh}}{\partial Q_k},\\
	\frac{dQ_k}{dt}  &=  \frac{\partial H_{B}^{Eh}}{\partial P_k},
	\end{align}	 
where 
\begin{equation}
	H_{B}^{Eh}(t) = \frac{1}{2} \sum_k \Big[ P_k^2 + \omega_k^2 Q_k + 2\alpha \bar{\sigma}_z(t) c_k Q_k \Big], 
	\end{equation}
and $\bar{\sigma}_z(t) = \mathrm{Tr}_S[\rho_S(t) \sigma_z]$

A second-order Runge-Kutta scheme was used to calculate $\mathcal{C}^{(ij)}_{nm}(t)$.  During individual time steps, $\bar{\sigma}_z(t)$ is kept constant for the evolution of the bath, while $\lambda^{cl}(t)$ is kept constant during the evolution of the system. Over a half time step, the equations for the classical variables take the forms,
	\begin{equation}
    \begin{split}
    Q_{k}\left(t + \frac{\delta t}{2} \right) &= \gamma_{k}(t)\cos\left( \frac{\omega_{k} \delta t}{2} \right) - \frac{\alpha c_{k}}{\omega_{k}^2}\bar{\sigma_z}(t)\\
    &\qquad \qquad + \frac{P_{k}(t)}{\omega_{k}}\cos\left(\frac{\omega_{k} \delta t}{2}\right) ,
    \end{split}
    \end{equation}
and 

    \begin{equation}
    \begin{split}
    P_{k}\left(t + \frac{\delta t}{2} \right) &= P_{k}(t)\cos\left(\frac{\omega_{k} \delta t}{2} \right) + \omega_{k}\gamma_{k}(t)\sin\left(\frac{\omega_{k} \delta t}{2}\right),
    \end{split}
    \end{equation}
where
    \begin{equation}
    \gamma_{k}(t) =  Q_{k}(t) + \frac{\alpha c_{k}}{\omega_{k}^2}\bar{\sigma_z}(t).
    \end{equation} 
Approximately $10^5$ trajectories were necessary to achieve convergence.

\end{document}